\documentclass[twocolumn]{aastex631}

\usepackage{amsmath}
\usepackage{makecell}
\usepackage{CJKutf8}
\newcommand{\cntext}[1]{\begin{CJK}{UTF8}{gbsn}#1\end{CJK}\kern-1ex}
\graphicspath{{./}{figures/}}
\usepackage{booktabs}

\submitjournal{ApJ}
\accepted{March 5, 2025}
\shortauthors{Wang M. et al.}

\begin{document}

\title{Two Phases of Particle Acceleration of a Solar Flare Associated with \textit{in situ} Energetic Particles}

\author[0000-0002-2633-3562]{Meiqi Wang (\cntext{王美祺})}
\affiliation{Center for Solar-Terrestrial Research, New Jersey Institute of Technology, 323 Martin Luther King Jr Blvd., Newark, NJ 07102-1982,
USA}

\author[0000-0002-0660-3350]{Bin Chen (\cntext{陈彬})}
\affiliation{Center for Solar-Terrestrial Research, New Jersey Institute of Technology, 323 Martin Luther King Jr Blvd., Newark, NJ 07102-1982,
USA}

\author[0000-0002-4795-7059]{Trevor Knuth}
\affiliation{NASA Goddard Space Flight Center, 8800 Greenbelt Rd, Greenbelt, MD 20771}

\author[0000-0002-0978-8127]{Christina Cohen}
\affiliation{Space Radiation Laboratory, California Institute of Technology, Pasadena, CA 91125, USA}

\author[0000-0002-5865-7924]{Jeongwoo Lee}
\affiliation{Center for Solar-Terrestrial Research, New Jersey Institute of Technology, 323 Martin Luther King Jr Blvd., Newark, NJ 07102-1982,
USA}

\author[0000-0002-5233-565X]{Haimin Wang}
\affiliation{Center for Solar-Terrestrial Research, New Jersey Institute of Technology, 323 Martin Luther King Jr Blvd., Newark, NJ 07102-1982,
USA}

\author[0000-0003-2872-2614]{Sijie Yu}
\affiliation{Center for Solar-Terrestrial Research, New Jersey Institute of Technology, 323 Martin Luther King Jr Blvd., Newark, NJ 07102-1982,
USA}
 
\begin{abstract}
How impulsive solar energetic particle (SEP) events are produced by magnetic-reconnection-driven processes during solar flares remains an outstanding question. Here we report a short-duration SEP event associated with an X-class eruptive flare on July 03, 2021, using a combination of remote sensing observations and \textit{in situ} measurements. The \textit{in situ} SEPs were recorded by multiple spacecraft including the Parker Solar Probe. 
The hard X-ray (HXR) light curve exhibits two impulsive periods. The first period is characterized by a single peak with a rapid rise and decay, while the second period features a more gradual HXR light curve with a harder spectrum. Such observation is consistent with \textit{in situ} measurements: the energetic electrons were first released during the early impulsive phase when the eruption was initiated. The more energetic \textit{in situ} electrons were released several minutes later during the second period of the impulsive phase when the eruption was well underway. 
This second period of energetic electron acceleration also coincides with the release of \textit{in situ} energetic protons and the onset of an interplanetary type III radio burst. We conclude that these multi-messenger observations favor a two-phase particle acceleration scenario: the first, less energetic electron population was produced during the initial reconnection that triggers the flare eruption, and the second, more energetic electron population was accelerated in the above-the-looptop region below a well-developed, large-scale reconnection current sheet induced by the eruption.

\end{abstract}


\section{Introduction} \label{sec:intro}
Solar energetic particle (SEP) events were first reported using \textit{in situ} observations in the 1960s \citep{vanAllen1965, Anderson1966}.
Some of these events that feature a short duration of a few hours are sometimes referred to as ``impulsive" SEP events \citep{Reames1999}. These events are usually accompanied by enrichment in the $^3$He isotope ($^3\mathrm{He}/^4\mathrm{He} \geq 0.01$; \citealt{Lin1996, Reames2021}) and have been argued to have a close association with reconnection processes in solar flares, although recent studies suggest that some of the $^3$He-rich SEP events are related to large-scale coronal EUV waves \citep{Nitta2015, Buvc2016}. 

The energetic electron component of SEPs is usually referred to as solar energetic electron (SEE) events.  
\citet{Wang2012} reported that nearly all of the SEEs were associated with type III radio bursts, which are a type of solar radio bursts driven by propagating electron beams along open field lines with a bulk speed of $\sim$0.1--0.5$c$ \citep{Wild1950, Reid2014, Chen2013, Chen2018}. 
These SEEs also have the characteristic of a beamed pitch-angle distribution and a time-of-flight velocity dispersion: electrons at higher energy arrive at the spacecraft first, followed by lower-energy ones later. The velocity dispersion can be generally modeled by electrons traveling along interplanetary field lines connecting the flare site and the spacecraft \citep{Lin1985, Krucker1999}, although the inferred path length sometimes differs significantly from that obtained from the Parker spiral configuration \citep{saiz2005, Kahler2006}.

Timing analysis of the release times of these \textit{in situ} electrons, compared with remote-sensing observations of their emission signatures, has offered important insights into their origins. The onset of impulsive electron events has been reported to have a good temporal relation with type III radio bursts and other flare signatures \citep{Kallenrode1991, Ergun1998, Nitta2008, Gomez-Herrero2021}, while delayed cases are also commonly reported \citep{Krucker1999, Haggerty2002, Cane2003, Wang2016}. For example, using observations made by the 3-D Plasma and Energetic Particles experiment on the WIND spacecraft (WIND/3DP; \citealt{Lin1995}) at a heliocentric distance of 1 AU, \citet{Krucker1999} reported that low-energy electrons (below 25 keV) were mostly associated with type III radio bursts, whereas electrons with high-energy (greater than 25 keV) were delayed by up to half an hour. They suggested that the delayed energetic electron component was more likely associated with acceleration by propagating Moreton waves. Likewise, \citet{Haggerty2002} attributed energetic electrons with a median delay of 10 minutes to those accelerated by coronal shocks associated with coronal mass ejections (CMEs). Some other studies \citep{Cane2003,Dresing2021}, however, argued that these electrons are instead injected simultaneously with their lower-energy counterpart responsible for the type III radio bursts, but the delays were attributed to transport effects in the interplanetary medium.

Furthermore, systematic energy-dependent delays of energetic electrons from the \textit{in situ} observations have been reported recently \citep{Li2020, Li2021, Wu2023}. \citet{Li2021} noted that these delays were caused by outward-propagating electrons undergoing a longer acceleration process than downward-propagating electrons, based on a comparison of the release times of \textit{in situ} electrons with HXR emitting electrons. \citet{Wu2023} found that energy-dependent delays commonly occurred in 26 out of 29 impulsive SEE events through statistical analysis. Meanwhile, these energy-dependent delays are widely reported using remote-sensing HXR observations of flare events alone \citep{Bai1979, Takakura1983, Qiu2004, Liu2015}. Possible scenarios, including repeated acceleration \citep{Lu1990, Liu2015} or two-step acceleration \citep{Bai1979, Qiu2004}, have been proposed to account for such delays.

Eruptive solar flares often show an extended period of X-ray and microwave bursts following the initial impulsive burst \citep{Kosugi1988,Bai1989}. For these events, the X-ray emission sometimes displays a soft-hard-harder (SHH) spectral evolution \citep{Cliver1986,Dennis1988}. A close connection between SEP events and flares with SHH behavior was established by \citet{Kiplinger1995} and subsequently confirmed by more recent studies \citep[e.g.,][]{Saldanha2008,Grayson2009,Kahler2012}. The extended X-ray/microwave emission is considered as the consequence of an extended energy release period and/or a delayed acceleration process \citep{Kosugi1983,Kai1986,Cliver1986}. Alternatively, they are interpreted as due to separate acceleration processes associated with a pattern of magnetic structural change due to reconnection \citep{Lee2018,Kliem2021}. A possible causal relation between the evolution of flare emission and SEP production is therefore important for understanding the full picture.

The energy spectra of SEEs also carry important information for deciphering their acceleration and/or transport processes. The \textit{in situ} electron peak flux typically displays broken power-law shapes \citep{Lin1982, Krucker2009}. \citet{Krucker2007a} reported that while the spectral indices of impulsive \textit{in situ} SEE electrons events show a positive correlation with those derived from the associated HXR bursts. Such a positive correlation has been confirmed by more recent studies \citep{Dresing2021, Wang2021}. Based on the observed spectral breaks, \citet{Wang2021} suggested that the source height of some SEEs in the corona was $\geq 1.3$ solar radii. More recently, a statistical study carried out by \citet{WangW2024} found that SEEs with different spectral types had different correlations with flares, CMEs, and CME-driven shocks. Intriguingly, all of these studies reached a similar conclusion that the total number of escaping electrons into interplanetary space was a very small fraction (0.1--1\%) of the number of HXR-producing electrons near the solar surface.

Here we investigate a SEP event associated with an X1.5-class solar flare on 2021 July 3 using joint \textit{in situ} measurements and remote-sensing observations in multiple wavelengths, including both X-rays and microwaves. The event displays an intriguing energy-dependent release of \textit{in situ} energetic electrons consistent with hardening HXR spectra, which we attribute to two distinct phases of particle acceleration processes during the evolution of the eruptive flare. In Section~\ref{observation}, we present an overview of the event. 
In Section~\ref{sec:timing}, we present our timing analysis based on the energy-dependent measurements of energetic electrons and protons, and compare the results to the remote-sensing observations. In Section~\ref{sec:spectra analysis}, we analyze the \textit{in situ} energetic electron spectra
observed by WIND and ACE and compare them to those derived from HXR observations. Finally, we summarize our main results in Section~\ref{Discussion} and discuss their implications.

\section{Observations} \label{observation}
\subsection{Event Overview}\label{sec:Event Overview}

\begin{figure}
    \center
    \includegraphics[width=0.4\textwidth]{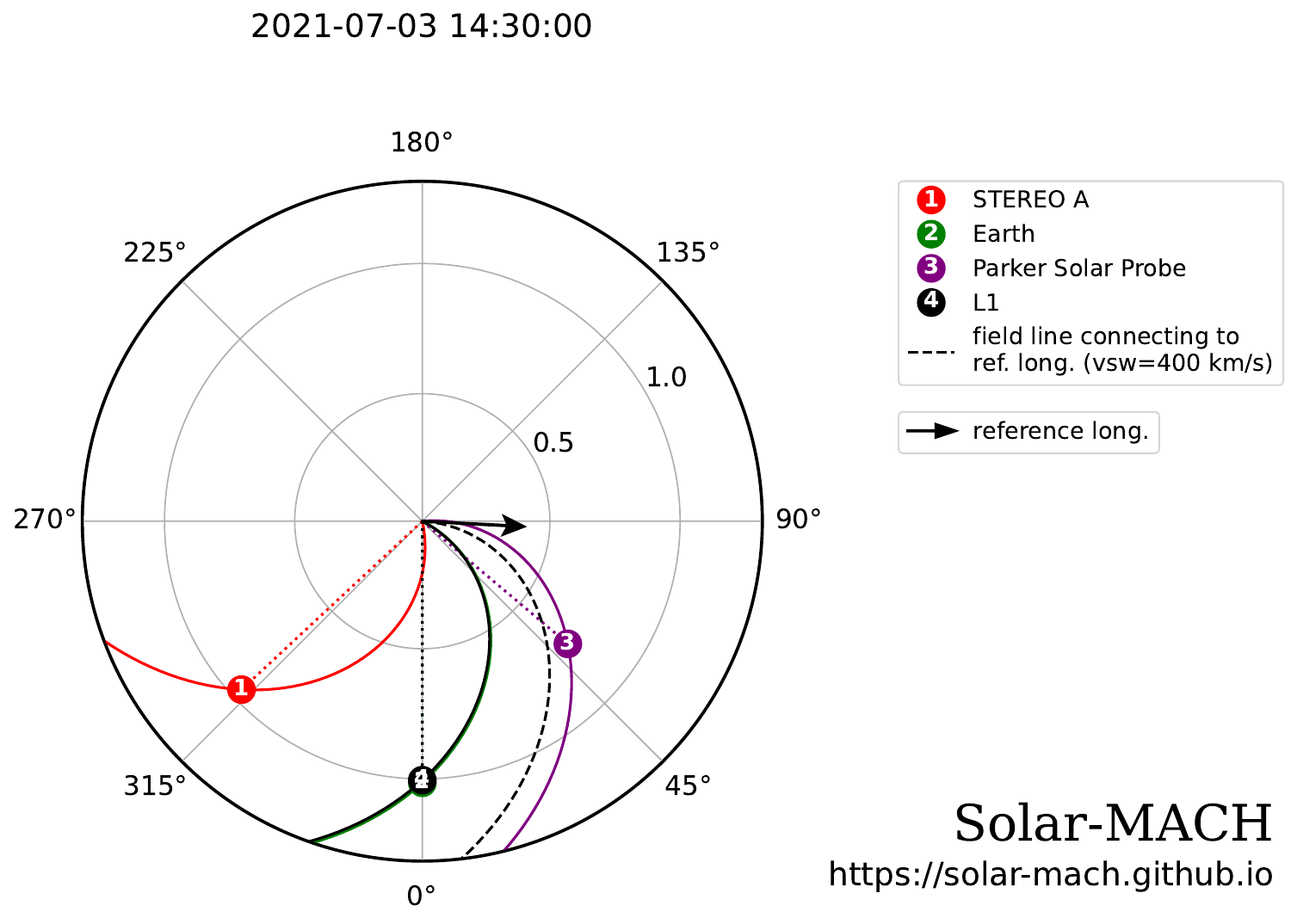}
    \caption{Locations of PSP, STEREO-A, and Earth's location at the time of the X class flare on 2021 July 3. The different curves in different colors correspond to the Parker spiral lines connecting the spacecraft to the solar surface. A solar wind speed is assumed to be 400 km~s$^{-1}$. The black arrow indicates the flare region and the black dashed curve indicates the Parker spiral line rooted at this region.}
    \label{Solar_mach}
\end{figure}

\begin{figure*}[!ht]
    \center
    \includegraphics[width=1\textwidth]{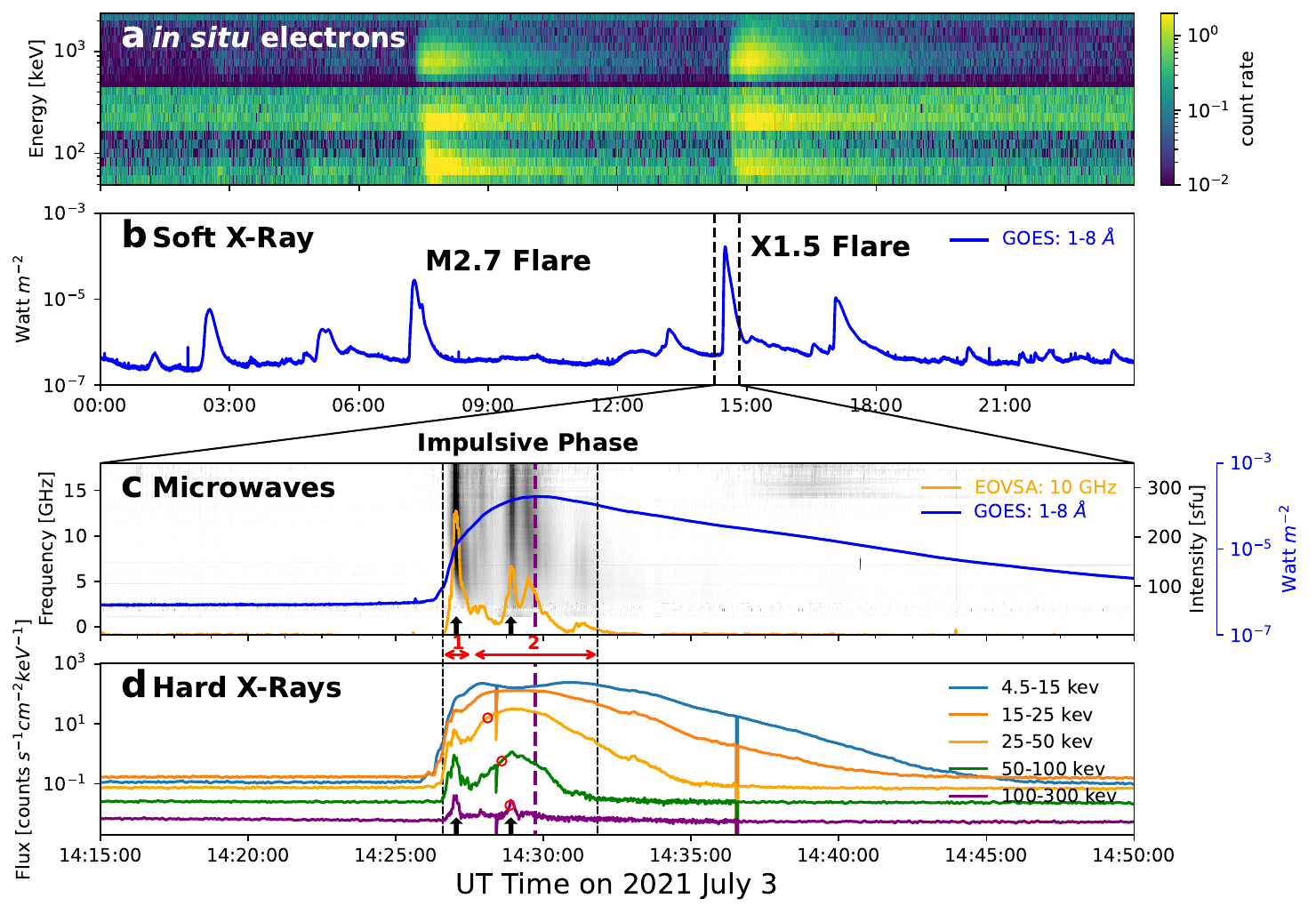}
    \caption{Time history of the 2021 July 3 event. (a) Electron count rate observed by PSP/IS$\odot$IS on July 03, 2021. (b) GOES 1--8 \AA\ soft X-ray light curve. The X1.5 flare at 14:29 UT is our focus of study. (c) Microwave dynamic spectrum, displayed with a grayscale color scale, was obtained by EOVSA from 14:15 to 14:50 UT for the X1.5 flare (time interval within the two dashed lines in (b)). The orange curve represents the light curve of EOVSA at 10 GHz. The two black arrows represent the first and second microwave peaks. 
    (d) HXR light curves observed by Fermi/GBM. The red double-sided arrows indicate the first and second impulsive periods during the main flare impulsive phase. The red circles indicate the half-maximum point of each energy channel during the rise phase of the second impulsive period. The two black arrows represent the first and second HXR peaks in the energy range of 100--300 keV. The two black vertical lines indicate the start and end of the two impulsive periods. The purple vertical dashed lines in (c) and (d) represent the onset time of the interplanetary type III radio burst.}
    \label{overview_obs}
\end{figure*}

On 2021 July 3, at around 15:00 UT, a short-duration SEP event was observed by the Integrated Science Investigation of the Sun (IS$\odot$IS; \citealt{McComas2016}) on board the Parker Solar Probe (PSP), which was located at a heliocentric distance of 0.74 AU. The SEP event was associated with an eruptive X1.5-class solar flare, which is the first X-class flare of Solar Cycle 25, from active region (AR) 12838 with its soft X-ray (SXR) flux peaking at 14:29 UT (SOL2021-07-03T14:29). Figure~\ref{overview_obs}(a) shows the electron count rate recorded by the two Energetic Particle Instruments onboard IS$\odot$IS (EPI-Lo and EPI-Hi) throughout the entire day.  In addition to the event of interest,  another short-duration SEE event was recorded at $\sim$07:30 UT, which corresponded to an earlier M2.7-class flare peaked at $\sim$07:17 UT. The two SEE events were also detected by WIND/3DP and the Electron, Proton, and Alpha Monitor (EPAM; \citealt{Gold1998}) instrument onboard the ACE spacecraft located near Earth (i.e., at a distance of 1 AU from the Sun). The locations of PSP, Earth, and STEREO-A (one of the Solar Terrestrial Relations Observatory, or STEREO, spacecraft; \citealt{Kaiser2008}), as well as their magnetic connectivities (assuming the ideal Archimedean Parker spiral with a solar wind speed of 400 km~s$^{-1}$), are presented in Figure~\ref{Solar_mach} (generated using Solar Mach, an open-source Python tool; \citealt{Gieseler2022}).  
Both of the two flares originated from the same AR 12838 located on the western limb. 
The close time correlation between the two SEE events observed by PSP and the corresponding flares, as well as the small longitudinal separation between the presumed magnetic footpoint of PSP and AR 12838 ($\approx 6^\circ$, marked as the black arrow in Figure~\ref{Solar_mach}) suggest that they should share the same origin. WIND/ACE spacecraft, whose magnetic footpoint was separated by approximately $27^\circ$ longitudinally from the flare region, also recorded SEE enhancements. STEREO-A did not record SEE events probably because of its large longitudinal separation from the flare region ($\approx77^\circ$).

Both the Expanded Owens Valley Solar Array (EOVSA; \citealt{Gary2018}) and the Gamma-ray Burst Monitor (GBM) aboard the Fermi observatory \citep{Meegan2009} had excellent coverage of the X1.5 flare in microwaves and HXRs, respectively. Figures~\ref{overview_obs}(c) and (d) show, respectively, the EOVSA microwave dynamic spectrum and Fermi/GBM HXR light curves at different energies. Multiple microwave and HXR peaks are present during the impulsive phase of the flare, which is defined as the period when HXR and microwave flux show a strong enhancement (see, e.g., \citealt{Benz2017}; roughly demarcated by the vertical dashed lines in Figures~\ref{overview_obs}(c)).
While the HXR light curves also feature multiple peaks during the impulsive phase, they have slightly different characteristics: 
During the initial impulsive phase, from 14:26:36 to 14:27:33 UT (marked by the red double-sided arrow labeled ``1'' in Figure~\ref{overview_obs}), both the microwave and HXR light curves are characterized by a single peak with a rapid rise and decay. 
After 14:27:41 UT (marked by the red double-sided arrow labeled ``2'' in Figure~\ref{overview_obs}), however, the HXR light curves are more gradual and have an increasingly longer duration at lower energies. 
In addition, the HXR flux exhibits an earlier rise at lower energies. Such an energy-dependent rise can be clearly seen in Figure~\ref{overview_obs}(d), where we mark the time when the flux reaches $1/2$ of its respective maximum (red circles) for each HXR light curve. For instance, the half-maximum point of the 25--50 keV channel occurs $\sim$45 s ahead of the 100--300 keV energy channel. 
We refer to the two distinct HXR peaks during the flare impulsive phase as the ``first impulsive period'' and the ``second impulsive period'' hereafter.

\subsection{EUV and Microwave Imaging} \label{sec:cite}

\begin{figure*}[!ht]
    \center
    \includegraphics[width=1\textwidth]{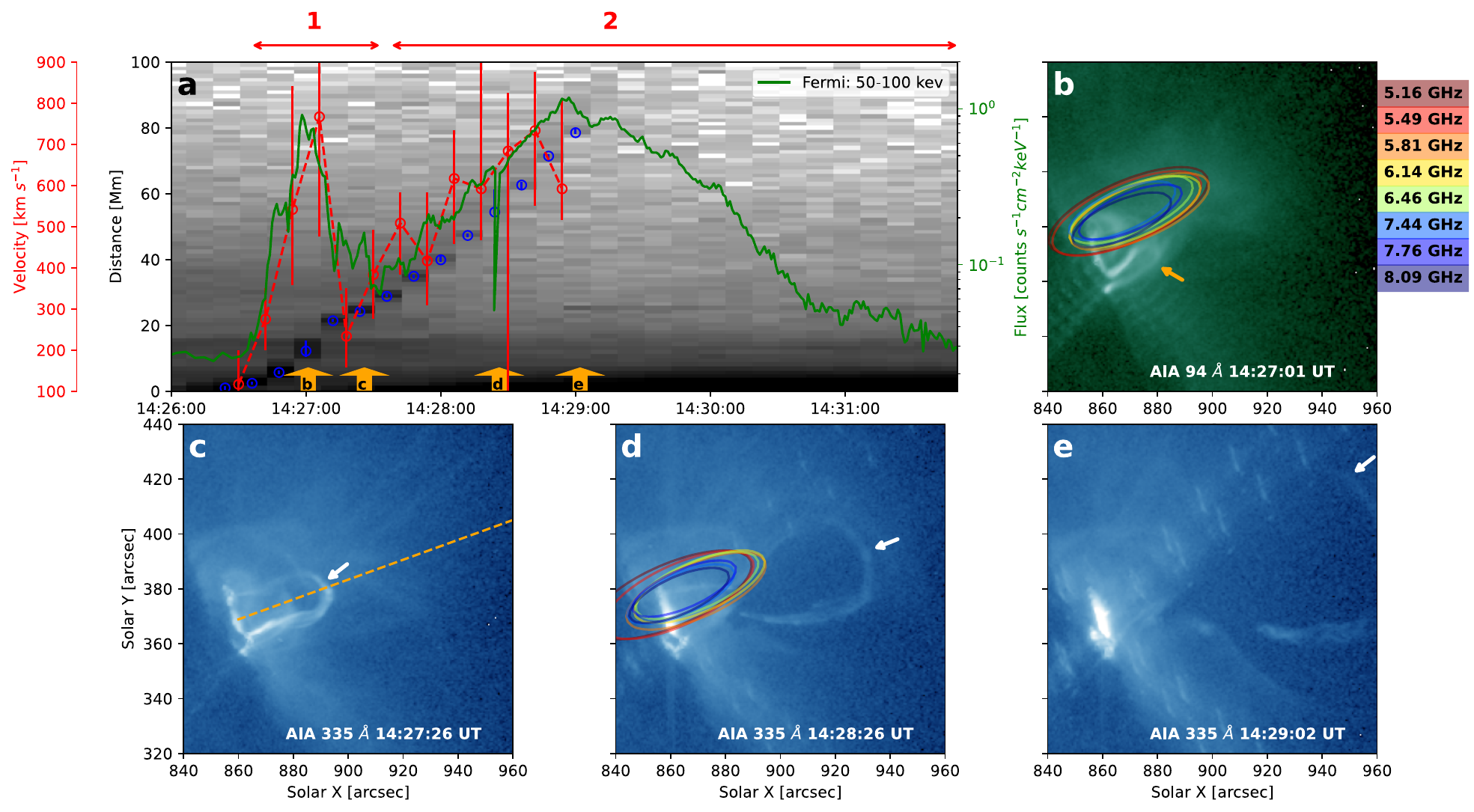}
    \caption{Flux rope eruption associated with the 2021 July 3 event. (a) Time-distance map derived from the slit shown as the orange dashed line in (c). The blue circles mark the maximum brightness on the map, representing the height of the erupting filament's leading front. The short vertical blue bars denote the height range enclosed by 80\% of the maximum brightness at each time, used as the error estimate. The red circles show the evolution of the speed of the erupting filament. The red vertical bars denote the estimated uncertainties. The green curve shows the 50-100 keV HXR light curve observed by Fermi/GBM. The times of SDO/AIA EUV images shown from (b) to (e) are indicated by the orange arrows in (a). Panels (b)--(e) illustrate the process of flux rope eruption seen by SDO/AIA 335 \AA. The EOVSA microwave sources at selected frequency channels are shown as open contours in blue to red colors corresponding to decreasing frequencies (90\% of the respective maximum). The orange arrow indicates the EUV brightening at the peak of the first impulsive period. The white arrows mark the erupting flux rope.}
    \label{flux_rope}
\end{figure*}

\begin{figure*}[!ht]
    \center
    \includegraphics[width=0.8\textwidth]{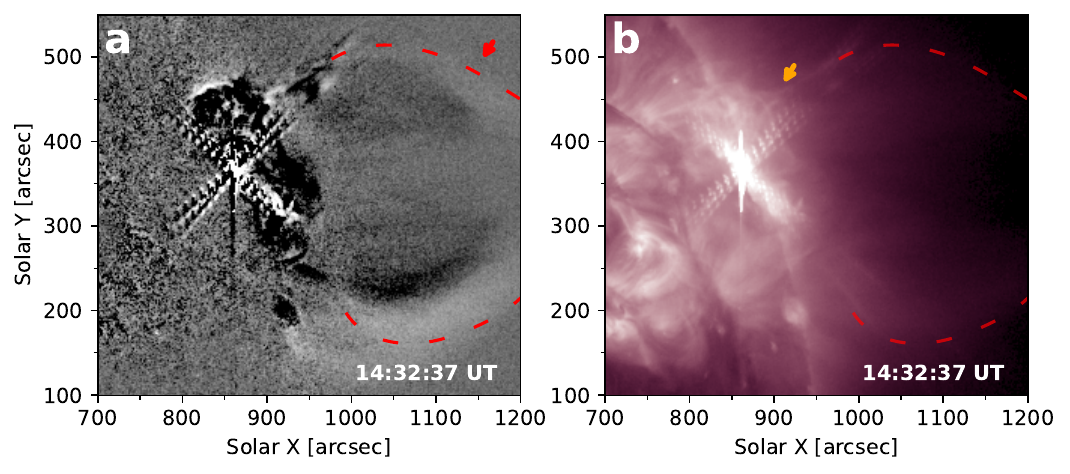}
    \caption{A possible signature of external magnetic reconnection during the flux rope eruption. (a) Difference map of SDO/AIA 211 \AA\ channel at 14:32:37 UT from 14:31:33 UT. The red arrow indicates the CME bubble driven by the erupting magnetic flux rope. (b) SDO/AIA 211 \AA\ image at 14:32:37 UT, which shows a possible signature of interchange reconnection between the erupting flux rope and the ambient open field lines. The orange arrow marks the potential magnetic reconnection site. The red dashed line in (a) and (b) outlines the approximate leading edge of the CME bubble at $\sim$14:32:37 UT derived from the difference map. The accompanying animation is available in the online journal. The animation presents SDO/AIA images in the 211 \AA\ channel, covering the time period from 14:26:45 UT to 14:34:45 UT with a 12-second cadence. The animation corresponds to panel (b) and aims to illustrate the temporal evolution of the external magnetic reconnection process during the flux rope eruption.}
    \label{interchange_reconnection_td}
\end{figure*}

Figure~\ref{flux_rope} shows time-series EUV images of the eruption observed by SDO/AIA. During the initial phase of the flare, a filament begins to erupt. During the first impulsive period (Figure~\ref{flux_rope}(b)), a newly brightened loop is seen by the AIA 94 \AA\ channel, as indicated by the orange arrow.
Starting around 14:27:26 UT, the filament, presumably a twisted magnetic flux rope, continued to erupt, forming a $\Omega$-like shape as evident in AIA 335 \AA\ channel(Figures~\ref{flux_rope}(c)--(e)). If the eruption conforms to the standard CSHKP eruptive flare model \citep{Carmichael1964, Sturrock1966, Hirayama1974, Kopp1976}, a large-scale current sheet can form below the erupting flux rope and drives energy release and particle acceleration (see, e.g., \citealt{Chen2020} and references therein).

To investigate the kinematics of the eruption, we produce a time-distance plot shown in Figure~\ref{flux_rope}(a), generated using a slit indicated by the orange dashed line in Figure~\ref{flux_rope}(c). The height of the leading edge of the erupting filament in the EUV images, projected in the plane of the sky, is marked by blue symbols. The speed of the erupting filament derived from adjacent height measurements is shown as red circles. The temporal behavior of the derived filament motion (red dashed curve) resembles that of the 50--100 keV HXR light curve (green curve) despite appreciable uncertainties. The flux rope undergoes two phases of acceleration. The first acceleration phase coincides with the first impulsive period, with the highest speed reaching $\sim$760 km~s$^{-1}$. During the second acceleration phase, the filament rises at a relatively lower speed with a more gradual increase in its speed. The eruption subsequently evolves into a white light CME recorded by the Large Angle Spectroscopic Coronagraph (LASCO; \citealt{Brueckner1995}) C2 and C3 from 2.69 to 19.84 solar radii, which propagates at a speed of $\sim$450 km~s$^{-1}$ (not shown here).

Figures~\ref{flux_rope}(b) and (d) show, as contours, the microwave sources at seven selected frequencies imaged during the first and second impulsive periods, respectively. Microwave imaging for this event is rather challenging for EOVSA, because the flare occurred in the early morning when the sun was only 8.5 degrees above the horizon, leading to a poor angular resolution in the east-west direction. 
Although the poor resolution\footnote{The synthesized beam size at the time of the observation is $132''\times47''$ at $\nu=5.16$ GHz and scales with $1/\nu$.} renders it difficult to pinpoint the exact microwave source location, during the second impulsive period, the source is likely located below the erupting filament close to the solar surface. 

During the eruption process after the second impulsive period, a brightening feature appears on the northern flank of the CME bubble enclosing the erupting magnetic flux rope (the red arrow in Figure~\ref{interchange_reconnection_td}(a)), as indicated by the orange arrow in Figure~\ref{interchange_reconnection_td}(b). We regard such a brightening as a possible signature of external magnetic reconnection induced by the interaction between the expanding CME bubble and the ambient open magnetic field lines. The reconnection creates a possible pathway for flare-accelerated energetic electrons to escape, as we will further discuss in Section~\ref{Discussion}.

\subsection{Low-Frequency Radio Bursts}\label{sec:type3}

\begin{figure*}[!ht]
    \center
    \includegraphics[width=0.8\textwidth]{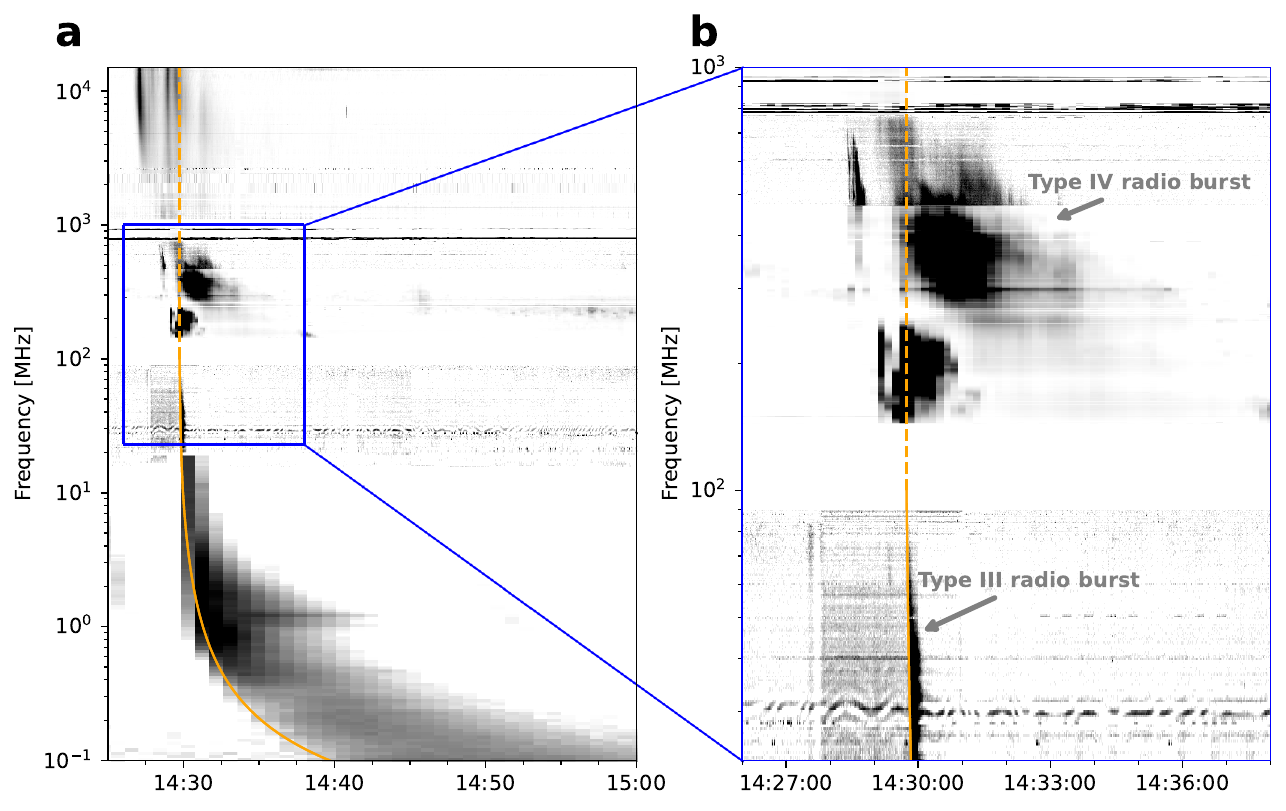}
    \caption{(a) Composite radio dynamic spectrum from 100 kHz to 18 GHz, recorded by EOVSA (1--18 GHz), e-Callistco/SWISS-BLEN7M (500--700 MHz), ORFEES/Nancay (150--470 MHz), e-Callistco/MRO (20--90 MHz), and PSP (0.1--20 MHz). The solid orange curve shows the fitted leading edge of the interplanetary type III radio burst observed by PSP and e-Callistco/MRO at $<$100 MHz. The vertical dashed orange line marks the derived onset time of the interplanetary type III radio burst at 14:29:44 UT. (b) Detailed view of the composite dynamic spectrum from the blue box in (a).}
    \label{type3}
\end{figure*}

At lower radio frequencies, multiple groups of solar radio bursts were recorded by ground-based radio stations around the world and space-based instruments onboard PSP, WIND, and STEREO-A.  
Figure~\ref{type3}(a) displays a composite dynamic spectrum covering the frequency range from 0.1 MHz to 18 GHz made by data from EOVSA (1--18 GHz), e-Callisto/SWISS-BLEN7M (500--700 MHz), Nan\c{c}ay/ORFEES (150--470 MHz),  e-Callisto/MRO (20--90 MHz), and PSP/FIELDS (0.1--20 MHz; \citealt{Bale2016}).

During the event, an interplanetary type III burst event was observed starting at around 14:30 UT. This burst event extends to metric-decametric wavelengths observed by e-Callistco/MRO and PSP, suggesting that the source electron beam entered the interplanetary space. In addition, a type IV radio burst, observed by e-Callisto/SWISS-BLEN7M in 500--700 MHz, started immediately following the type III radio burst. The burst drifts from high to low frequencies with a frequency drift rate of $\sim$2 MHz~s$^{-1}$, consistent with moving type IV radio bursts classified by a drift rate $\geq$0.03 MHz~s$^{-1}$ reported in the literature \citep{Robinson1978, Gergely1986,Kumari2021}, which were interpreted as radio emission associated with erupting materials. Figure~\ref{type3}(b) shows the detailed structure of the type III and IV bursts in the zoomed-in view of the blue box in (a).

We fit the leading edge of the type III radio burst in the dynamic spectrum from $\sim$100 MHz to 100 kHz using a similar method described in \citet{wang2023}, which adopted a combined coronal and solar wind density model with the electron beam speed $v_b$ and the release time of the electron beam $t_0$ as free parameters. Considering a possible decrease in the speed of the exciter beam as it propagates in the interplanetary medium, we also apply a polynomial model described in \citet{Krupar2015} (see their Equation 3), with the constant acceleration $a$ included as the third free parameter in the fitting. The fitted frequency drift curve is shown as the orange solid curve in Figure~\ref{type3}(a). The best-fit initial electron beam speed $v_b$ is 0.28c, which corresponds to a bulk kinetic energy of $\sim$20 keV. The best-fit acceleration $a$ is -15 $km/s^{2}$. The extrapolated onset time of the type III burst is 14:29:44 UT (shown as the vertical orange dashed line in Figure~\ref{type3}). It coincides with the second impulsive period, suggesting a possible correspondence with the microwave-emitting energetic electrons during the course of the filament eruption. 

\section{Release Times of the \textit{in situ} SEPs} \label{sec:timing}

\subsection{Energetic Protons} \label{proton vda}

\begin{figure}
    \center
    \includegraphics[width=0.5\textwidth]{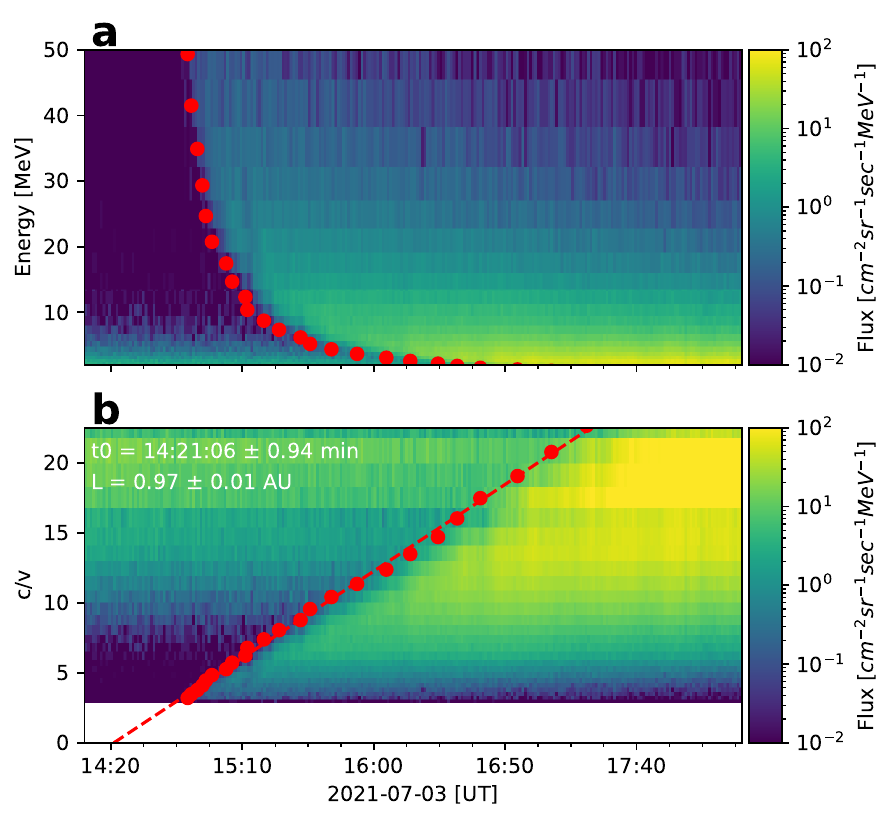}
    \caption{(a) Proton flux observed by PSP/IS$\odot$IS at different energies. The red circles correspond to the onset times of the proton flux at each energy channel. (b) Same data as (a), but the vertical axis is transformed into $c/v$ to display the velocity dispersion. The red dashed line shows the linear fit results of the energy-dependent onset times.}
    \label{vel_dis_p}
\end{figure}

Starting from $\sim$14:47 UT, IS$\odot$IS on board PSP recorded an enhancement of the energetic proton flux in 0.7--50 MeV, as shown in Figure \ref{vel_dis_p}(a). To determine the release time of the energetic protons, we adopt a method of fitting the logarithm of the protons' early rise phase, similar to that of \citet{Miteva2014}. 
The derived energy-dependent onset times are shown as the red circles in Figure~\ref{vel_dis_p}. It is evident that a systematic delay of the onset time toward lower energies is present. 

In Figure~\ref{vel_dis_p}(b), the same data are shown, but the proton energy is now transformed into $c/v$, where $c$ represents the speed of light and $v$ is the speed of protons. With this representation, a nearly linear relation between the onset time and $c/v$, known as the ``velocity dispersion'' due to the differences in transit time from the particle source to the spacecraft for different energy particles, can be clearly seen. Such a relation is only present if the particles are released from their source region simultaneously. They traverse the interplanetary space and reach the spacecraft with a similar path length of $L$.  
In this case, the release time of the particles at the solar source $t_0$ can be extrapolated using the linear relation $1/v = (t - t_0)/L$ in the limit of $v\rightarrow \infty$ (or $c/v\rightarrow 0$), where $t_i$ represents the onset time measured at an energy channel that corresponds to a speed of $v_i$. The extrapolation result is shown as the red dashed line in Figure~\ref{vel_dis_p}(b), with the release time $t_0$ (intersection of the line with the $x$ axis) found to be at 14:21:06 UT with an uncertainty of 0.94 minutes. 
The derived path length $L\approx0.97$ AU is $\sim$22\% longer than the nominal Parker spiral path length of 0.81 AU from the solar surface to PSP assuming a solar wind speed of 400 km~s$^{-1}$. 

\subsection{Energetic Electrons} \label{electron vda}

\begin{figure*}
    \center
    \includegraphics[width=0.9\textwidth]{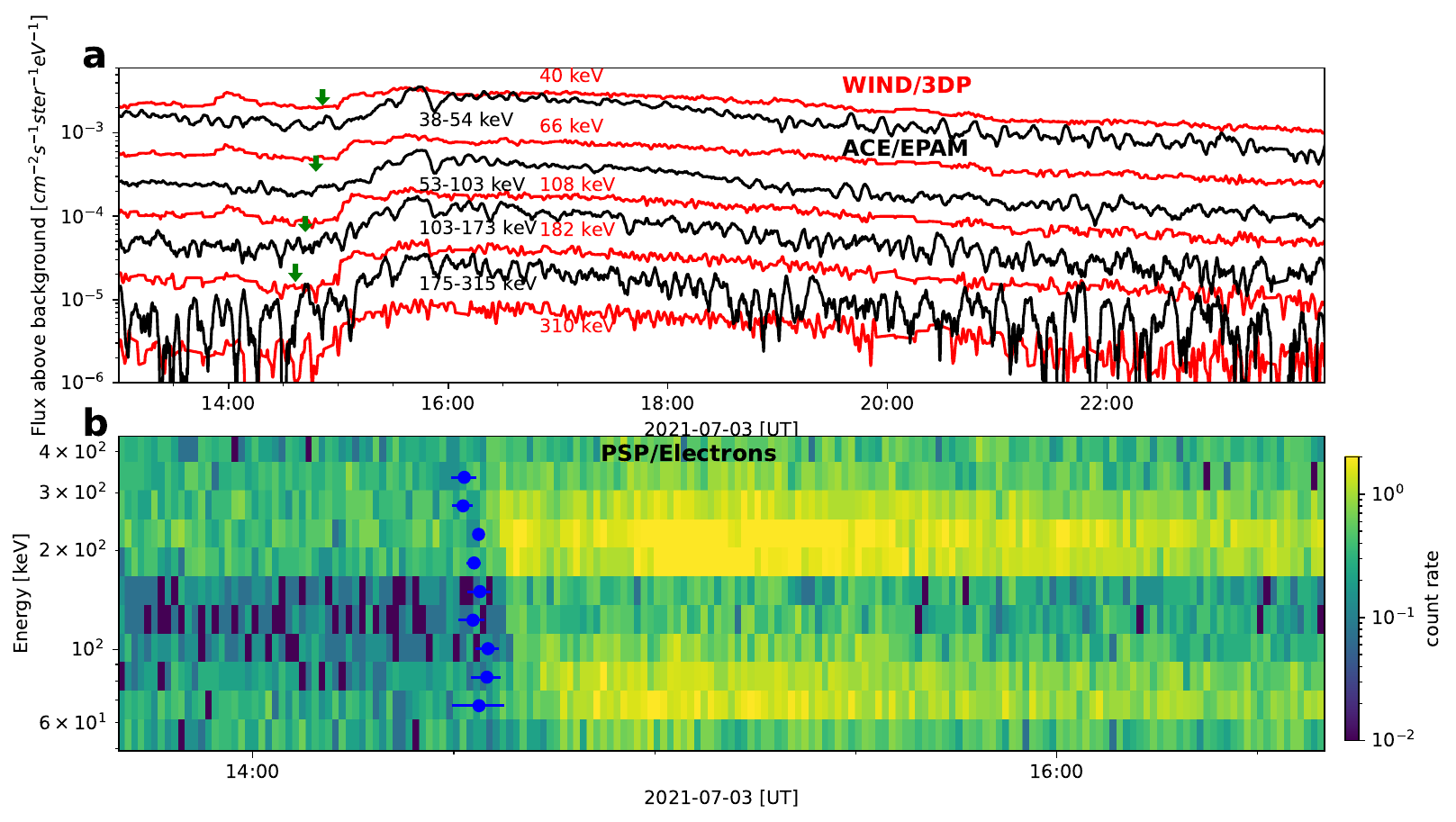}
    \caption{Energetic electrons observed by WIND, ACE, and PSP. (a) Differential \textit{in situ} electron flux recorded by WIND/3DP (red) and ACE/EPAM (black). The flux has been pre-event-background-subtracted and smoothed. The green arrows indicate the onset times at different energies. (b) Electron count rates observed by PSP/IS$\odot$IS. The blue points with error bars represent the onset time of each energy channel analyzed.}
    \label{vel_dis_e}
\end{figure*}

For energetic electrons associated with this event, WIND/3DP and ACE/EPAM both detected an enhancement in the differential flux at 1 AU, shown in Figure \ref{vel_dis_e}(a). The four energy channels, namely DE1 (38--54 keV), DE2 (53--103 keV), DE3 (103-173 keV), and DE4 (175--315 keV), recorded by ACE/EPAM, observed an enhanced energetic electron flux, represented as the black curves in Figure~\ref{vel_dis_e}(a). 

During the same period, five energy channels from 40 keV to 310 keV, measured by WIND/3DP, exhibit electron flux profiles similar to those observed by ACE/EPAM (red curves in Figure~\ref{vel_dis_e}(a)), suggesting that they share a common source as those observed by ACE. All the electron flux profiles have been background subtracted and smoothed using the Savitzky-Golay method \citep{Savitzky1964}.  In this case, the rise phase of the WIND/ACE flux has large fluctuations, and a simple exponential shape might not provide the best fit. Here we define the onset time of the \textit{in situ} energetic electron event as the time when the electron flux reaches three times the root mean square fluctuation of the pre-event background flux ($3\sigma$). The derived onset times at the ACE spacecraft are marked as the green arrows in Figure~\ref{vel_dis_e}(a). 
Similar to the energetic protons, they display a clear delay toward lower electron energies. We further utilize the velocity dispersion method to determine the release time of these electrons, which returns a release time of 14:18:04 UT with an uncertainty of 4.4 minutes and a propagation path length of $1.7\pm0.3$ AU.At 0.74 AU, EPI/Lo onboard PSP/IS$\odot$IS also detected the \textit{in situ} energetic electron event from $\sim$50--400 keV, shown in Figure~\ref{vel_dis_e}(c). We adopt a method of fitting the logarithm of the electrons' early rise phase, identical to that used for protons, to derive the onset time of the electrons, represented by blue circles. To estimate the uncertainties of the onset times, following \citet{Gehrels1986}, we use the confidence level of 0.90 based on Poisson statistics to estimate their upper and lower limits, which is more suitable for weaker events with fewer counts. To estimate the release time of these energetic electrons, we assume that they travel along the same path length as protons of 0.97 AU.

\begin{figure*}[!ht]
    \center
    \includegraphics[width=1\textwidth]{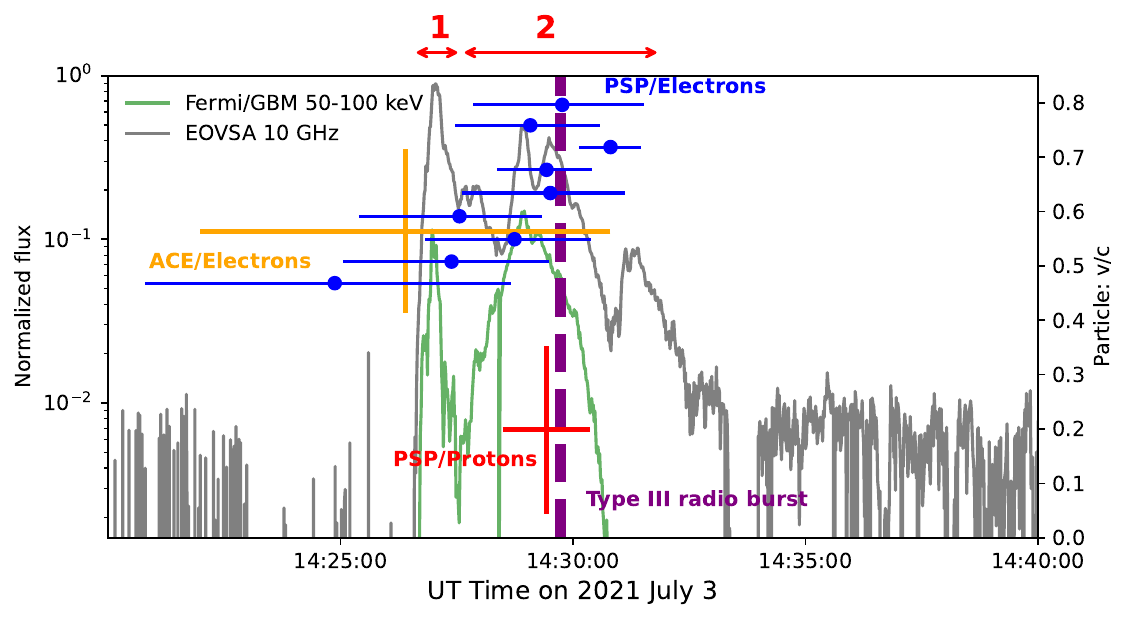}
    \caption{Comparison of the derived release times of \textit{in situ} particles with radio and HXR flare emissions. To enable a direct comparison, the light traveling time to the spacecraft location has been added to the particle release times. Gray and green curves represent EOVSA 10 GHz microwave and Fermi 50-100 keV HXR light curves. The flux has been background-subtracted and normalized from 0 to 1. The orange and red crosses, respectively, denote the derived release times of the $\sim$50--300 keV electrons measured by ACE and 2--50 MeV protons observed by PSP, respectively, as a function of their speeds (right vertical axis). The blue circles with error bars show the derived release time from PSP/electrons at the energy range of $\sim$60-400 keV. The double-sided arrows represent the first and second impulsive periods, respectively. The purple vertical dashed line represents the derived onset time of the interplanetary type III radio burst. }
    \label{release_time}
\end{figure*}

\subsection{Comparison with Radio/X-ray Emissions} \label{rel t summ}
In Figure~\ref{release_time}, we compare the derived onset times of the \textit{in-situ} SEP protons and electrons to remote-sensing radio and X-ray light curves. To compensate for the light traveling time from the Sun to Earth, we have added the light traveling time to the respective spacecraft location for all the \textit{in situ} particle release times. 
The release times of the electrons observed by both PSP and ACE are generally consistent with each other. Despite their relatively large uncertainties due to the low-cadence data available, the release times of the electrons observed by the PSP exhibit an apparent energy-dependent trend: the lower energy electrons (70--150 keV) are released earlier, starting around the peak of the first impulsive period, while the electrons at higher energy ($>$150 keV) are released at later times, which generally coincide with the second impulsive period. The energetic protons observed by PSP (red crosses) also appear to be released during the second impulsive period.

\section{Electron spectra
at the Sun and \textit{in situ}} \label{sec:spectra analysis}

\begin{figure}
    \center
    \includegraphics[width=0.4\textwidth]{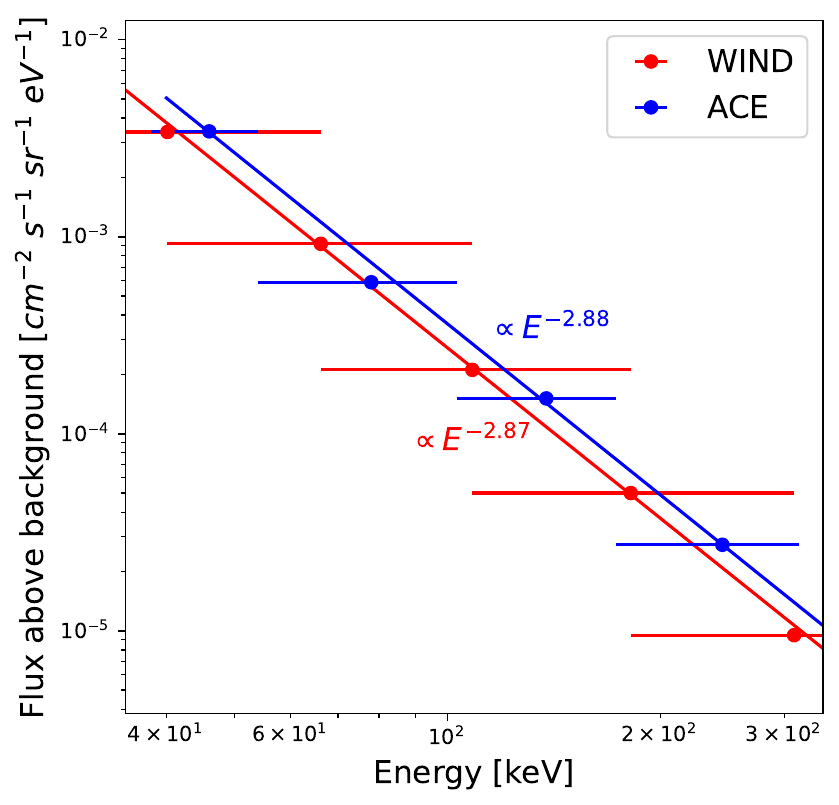}
    \caption{The background-subtracted \textit{in situ} energetic electron flux spectra obtained by the WIND (red) and ACE (blue) spacecraft. The blue and red lines are the results of power-law fitting. The horizontal bars represent the energy bin width for WIND and ACE.}
    \label{spectra_insitu}
\end{figure}

\begin{figure*}[!ht]
    \center
    \includegraphics[width=1\textwidth]{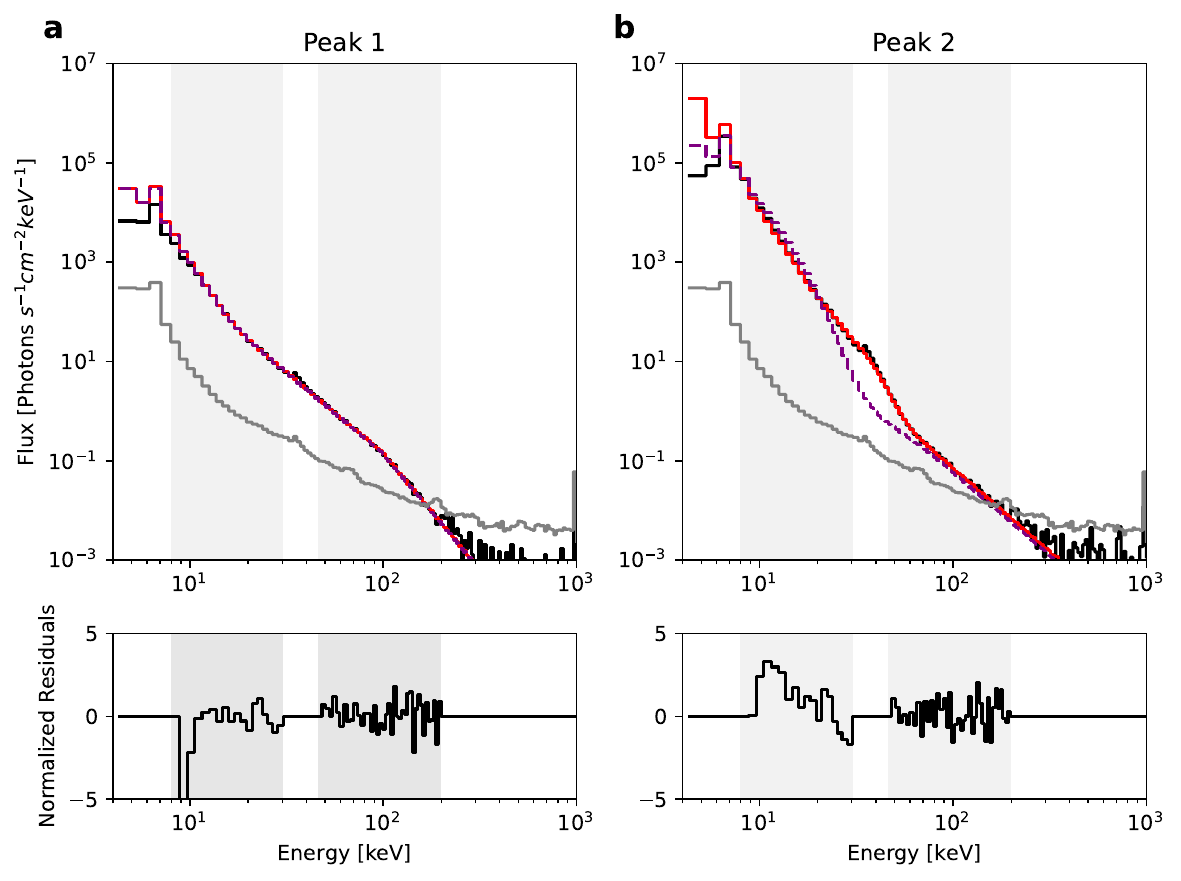}
    \caption{Fermi/GBM observed X-ray spectra (black curve) along with spectral fit results (red curve) for the first (a) and second (b) HXR peaks, respectively. Background subtraction has been applied. In each panel, the fit spectrum is represented by the red solid curve, which is a combination of an isothermal model and a broken power-law model for the non-thermal portion, with contributions from the estimated pulse pileup effects.  The purple dashed curve represents the output spectrum directly from the combined source model using the fit parameters without applying the pulse pileup correction output. The gray curve shows the background. The residuals are shown in the bottom panels. The best-fit parameters are shown in Table~\ref{para_tab}.}
    \label{spectra_hardxray}
\end{figure*}

\begin{figure}
    \center
    \includegraphics[width=0.4\textwidth]{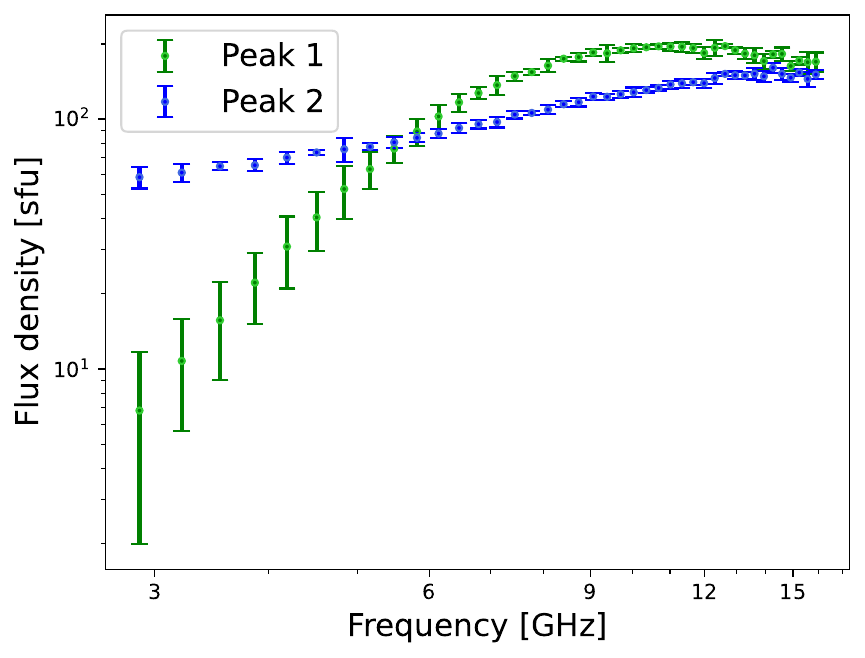}
    \caption{Total-power microwave spectra obtained by EOVSA for the first (green) and second (blue) microwave peaks.}
    \label{spectra_mw}
\end{figure}
\begin{table*}[!ht]
    \center
    \begin{tabular}{lllllll}
    \toprule
    ~ & EM  [$10^{48} cm^{-3}$] & $T$ [$10^{7} K$] & ${\varepsilon}_b$ [keV] & $\gamma_{l}$ & $\gamma_{h}$ \\ 
    \midrule
    HXR: $t_1$ & 1.11 & 2.20 & 97.62 & 3.26 & 4.65 \\  
    HXR: $t_2$ & 4.75 & 3.30 &  77.54 & 2.79 & 3.33 \\

    \bottomrule
    \end{tabular}
    \caption{Summary of spectral analysis results based on Fermi/GBM HXR data. $t_1$ and $t_2$ correspond to the first and second HXR peaks, respectively. EM and $T$ are the emission measure and temperature of the thermal plasma. $\gamma_{l}$ and $\gamma_{h}$ are the power-law spectral index below and above the break energy ${\varepsilon}_b$ of the HXR photon flux spectra. }
    \label{para_tab}
\end{table*}

We further explore the spectral properties of the energetic electrons measured in both the interplanetary space by the \textit{in situ} instruments and those near their solar source by utilizing HXR and microwave spectral analysis.

First, for the \textit{in situ} energetic electrons, we perform a power-law fitting method on the electron spectra measured by ACE/EPAM and WIND/3DP in Figure \ref{vel_dis_e}\footnote{PSP measurements of the energetic electrons are in raw count rates. Those in calibrated flux unit are unavailable at the time of this work, therefore are not used for spectral analysis.}. The fitted results, shown in Figure \ref{spectra_insitu}(a), return power-law indices of $\delta^{\rm ACE}=2.88\pm 0.17$ and $\delta^{\rm WIND}=2.87\pm 0.04$ derived from ACE and WIND data, respectively. 

Then, we perform spectral analysis on the two HXR peaks using the \texttt{OSPEX} package \citep{Schwartz2002} available in the \texttt{SSWIDL} distribution. The background-substracted spectra have been fitted with a combination of an isothermal model, which dominates the X-ray flux lower energy range, and a broken power-law photon distribution for the high-energy, presumably non-thermal component. Specifically, this broken power-law distribution function is parameterized by a power-law form with a spectral index $\gamma_{l}$ for energies below break energy ${\varepsilon}_b$ and transitions to a different power-law form with an index $\gamma_{h}$ for energies above ${\varepsilon}_b$.

For HXR spectra formed during strong solar flares, a pile-up effect can be significant. We select the data obtained from the detector toward less sunward and apply the pulse pile-up correction (PPU) technique as described by \citet{Lesage2023}. This technique takes into account the pulse pile-up effect, which occurs when two photons arriving at the detector at the same time are measured as a single photon with the summed energy of the two individual photons. 
The energy range used for the fitting is 8--200 keV. The 30-40 keV range is excluded from the fitting because the Fermi/GBM NaI detectors have an Iodine feature, which is not sufficiently captured in the instrument response matrix. 
Figures~\ref{spectra_hardxray}(a) and (b) display the best-fit X-ray spectra for the two HXR peaks marked as black arrows in Figure~\ref{overview_obs}(d). The corresponding fit parameters are listed in Table~\ref{para_tab}. The best-fit spectra with and without the PPU correction applied are shown by the red solid and purple dashed lines in Figures~\ref{spectra_hardxray}, respectively. It can be seen that the pileup effect is negligible in the first peak (panel (a)) but significant in the second peak (panel (b)) for the 20--60 keV energy range (which results from the 10--30 keV photons around the peak of the observed HXR photon count spectrum). 
For the first HXR peak, the derived photon spectral indices are $\gamma_{l} = 3.26$, $\gamma_{h} = 4.65$, separated at a break energy of ${\varepsilon}_b=97.6$ keV. For the second peak, the fitting parameters are $\gamma_{l} = 2.79$, $\gamma_{h} = 3.33$, and ${\varepsilon}_b=77.5$ keV. 
Although the low-energy portion ($<$60 keV) of the HXR spectra for the second HXR peak is strongly affected by the pulse pileup effect, the high-energy portion ($\gtrsim$100 keV) of the HXR spectra for the second peak is distinctively harder than that of the first HXR peak. If the high-energy HXR emission (above $\varepsilon_b$) falls into the thick-target bremsstrahlung regime, the corresponding spectral indices of the nonthermal electron distribution are $\delta_1 = 5.65$ and $\delta_2=4.33$ for the first and second HXR peak, respectively ($\delta\approx\gamma_h+1$).

In addition to the X-ray diagnostics, gyrosynchrotron radiation from $\gtrsim$100 keV energetic electrons complements the HXR diagnostics. The observed microwave total-power spectra at the two peaks (marked as the black arrows in Figure~\ref{overview_obs}(c)) are displayed in Figures~\ref{spectra_mw}. The observed microwave spectrum at the first time peak (green) exhibits a positive slope below a peak frequency of 14 GHz and a negative slope above it, which conforms to the characteristics of non-thermal gyrosynchrotron radiation \citep{Dulk1985}. The second microwave peak has a positive spectral slope across the entire 3--18 GHz range, indicating that the spectral peak could be located at $>$18 GHz. 
Due to the lack of frequency samplings of the spectra in the optically thin regime (i.e., frequencies with a negative spectral slope), as well as the unavailability of spatially resolved brightness temperature measurements due to unfavorable observing conditions for imaging, microwave spectral fitting is deemed difficult and might result in large uncertainties. Therefore, we do not attempt to perform quantitative microwave spectral analysis based on these data. On the other hand, since the peak frequency depends strongly on the magnetic field strength and the hardness of the energetic electron distribution (see, e.g., Equation 39 in \citealt{Dulk1985}), a hardening of the electron distribution, as suggested by HXR spectral analysis, is consistent with the observed shift of the peak frequency to higher frequencies from the first to the second impulsive period. 

With both \textit{in situ} measurements and remote-sensing HXR observations, we can further estimate the total number of electrons that escape to the interplanetary space and those retained near the solar surface, respectively. If the HXR emission falls into the thick-target regime, the total HXR–emitting electron rate $\dot{N_e}$ above 100 keV can be estimated from the observed photon flux spectrum assuming a power-law electron distribution (see, e.g., Eq. 2.12 in \citealt{Holman2011}). 
The total number of electrons $N_e\approx \dot{N_e}\tau$ where $\tau$ is the duration of the electron injection time estimated using the FWHM of the 100--300 keV HXR light curve. The estimates of the total number of electrons above 100~keV for the first and second HXR peaks are $4.04\times 10^{34}$ and $5.65\times 10^{34}$, respectively. 
We then estimate the number of escaping electrons from WIND and ACE measurements. Following \citet{James2017}, the total number of escaping electrons is calculated by integrating the electron distribution over energy (above the same selected energy of $E_0=100$ keV), the angular spread of electrons at 1 AU (taken as $40^{\circ}$), and the time duration of the electron event. The calculated total number of electrons above 100~keV using the WIND and ACE observations are $\sim$ $1.3\times 10^{31}$ and $\sim 1.7\times 10^{31}$, respectively. The ratio of escaping electrons to HXR-emitting electrons above 100~keV is approximately 0.03\% to 0.04\% for the first HXR peak and about 0.02\% for the second HXR peak.

\section{Discussion and Conclusion} \label{Discussion}

\begin{figure*}[!ht]
    \center
    \includegraphics[width=1\textwidth]{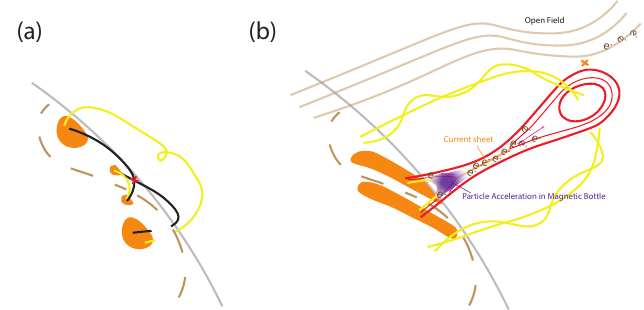}
    \caption{Schematic diagram of magnetic reconnection and the associated particle acceleration at different stages of the eruptive solar flare. (a) The first impulsive period may be associated with the initialization of the magnetic flux rope eruption. The ``tether-cutting'' scenario, in which a pair of highly sheared loops (black curves) undergo reconnection and form a twist flux rope (upper yellow curve), is depicted here as one of the possibilities. The brown dashed line represents the polarity inversion line.
    (b) During the second impulsive period, the scenario is similar to the standard CSHKP eruptive flare model. A large-scale reconnection current sheet formed behind the erupting flux rope, driving magnetic energy release. Charged particles may undergo prolonged acceleration to high energies in a ``magnetic bottle" structure located at the bottom of the current sheet. Possible reconnection between the magnetic field lines enclosing the erupting flux rope and the ambient open field line may provide a ``pathway'' for the flare-accelerated particles to escape into the interplanetary space.}
    \label{cartoon}
\end{figure*}

We performed a comprehensive study of a SEP event associated with an eruptive X flare observed by multiple ground-based and space-borne instruments by combining both remote sensing and \textit{in situ} measurements. We compare the multi-messenger observations in terms of their timing and spectral characteristics. We summarize our main findings below:

\begin{itemize}
\item The \textit{in situ} SEP event has a short duration ($\sim$3 h for WIND/ACE and $\sim$0.5 h for PSP for energetic electrons, and $\sim$3 h for energetic protons) with a rapid-rise, rapid-decay time evolution, a feature usually found to be associated with acceleration by flare reconnection \citep{Reames1999}.

\item The microwave and HXR flare emissions exhibit two distinct periods of bursts during the main flare impulsive phase. The first period is very impulsive, with a quick rise and decay, while the second period is more gradual in HXRs. We refer to the two periods as the first and second impulsive periods, respectively.

\item Timing analysis of the \textit{in situ} energetic electrons observed by ACE and PSP suggest that they are released during the flare impulsive phase. The release times of electrons observed by PSP seem to exhibit an energy-dependent delay, with lower-energy electrons being released earlier during the first impulsive period and higher-energy electrons being released later during the second impulsive period.

\item The \textit{in situ} energetic protons display a clear velocity dispersion in energy, suggesting a common release time associated with the second impulsive period. The derived path length is $\sim$22\% longer than the nominal Parker spiral length.

\item Spectral analysis indicates that the spectral indices derived from the \textit{in situ} energetic electrons and HXR-emitting electrons near the flare source are generally consistent with those reported previously. Notably, the spectral index of the high-energy electrons derived from the HXR data is harder during the second flare impulsive period. The fraction of escaping electrons measured \textit{in situ} to HXR-emitting electrons is extremely small, which is at the order of $\sim$0.01--0.1\%.
\end{itemize}

Here, we attempt to interpret the multi-messenger observations in a physical scenario within the framework of a two-phase particle acceleration process driven by flare reconnection. In our scenario, the eruptive flare undergoes two main phases of energy release, which are represented by the first and second HXR/microwave burst periods during the flare impulsive phase. During the first impulsive period, there is a noticeable EUV brightening, followed by the appearance of brightened loops (Figure~\ref{flux_rope}). Meanwhile, the leading front of the flux rope has an abrupt acceleration. The observed phenomena during this period are consistent with the initialization of large eruptive flares due to either a ``tether-cutting'' reconnection \citep{Moore2001,Liu2013, Jiang2021} or a ``breakout'' reconnection scenario \citep{Antiochos1999, Lynch2008, Chen2016}, which we cannot distinguish due to the lack of information on the magnetic field configuration of this limb event. 
After that, the erupting flux rope (seen in Figure~\ref{flux_rope}(d) as an $\Omega$-shaped structure) induces a large-scale current sheet above the flare arcade in a way similar to that depicted in the standard eruptive flare model, serving as the site for more prolonged magnetic energy release and particle acceleration. This process corresponds to the second HXR/microwave impulsive period, which features a more gradual HXR light curve and multiple microwave peaks.

We depict our scenario for the first and second impulsive periods in Figures~\ref{cartoon}(a) and (b), respectively. During the first impulsive period when the flux rope eruption is initiated, for either the tether-cutting or breakout reconnection scenario, the magnetic energy release likely occurs in a relatively compact, localized reconnection region. Figure~\ref{cartoon}(a) illustrates a possible geometry using the tether-cutting scenario as an example. As a result, the energy release likely proceeds in an impulsive manner, resulting in HXR/microwave light curves with rapid rise and decay features. Meanwhile, due to the impulsiveness of the associated particle acceleration, particles of different energies are accelerated and released at nearly the same time, giving rise to the clear velocity dispersion of the \textit{in situ} energies electrons observed by the ACE spacecraft at relatively low energies. The absence of more energetic electrons and energetic protons released during this period suggests that the particle acceleration process may not be efficient enough (e.g., in the presence of a large guide field; see, e.g., \citealt{Dahlin2016, Arnold2021}) and/or too short to accelerate a sufficient amount of high-energy electrons and protons to above detectable levels. 

With the presence of a well-developed, large-scale current sheet trailing the erupting flux rope, the reconnection geometry during the second impulsive period is very different. The picture of our perceived energy release and particle acceleration is illustrated in Figure~\ref{cartoon}(b). First, the large-scale nature of the magnetic reconnection region offers more space and free energy available for accelerating electrons and ions to high energies. Second, the often observed strong-to-weak shear evolution \citep[see, e.g.,][]{Aschwanden2001, Aulanier2012} during the flare energy release provides a more favorable condition for more efficient particle acceleration with a reduced guide field. Last but not least, a ``magnetic bottle'' structure naturally develops at the bottom of the current sheet \citep{Chen2020,Chen2024}, providing a potential site for trapping and (re-)accelerate particles to higher energies for a longer period. Mechanisms that could be responsible for such prolonged particle acceleration include fast-mode termination shocks \citep{Tsuneta1998, Guo2012, ChenB2015}, magnetic islands \citep{Drake2006,WangY2021,Guidoni2022}, collapsing traps \citep{Somov1997,Karlicky2004}, and turbulence/waves \citep{Kontar2017,Bacchini2024}. All these conditions favor the acceleration of electrons and ions toward higher energies during this period, which is supported by the hardening electron spectra derived from HXR spectral analysis. Likewise, the apparent energy-dependent delay of the $>$150 keV \textit{in situ} electrons, as derived from the PSP observations by assuming a common propagation length, may also be attributed to the more prolonged and possibly multi-staged acceleration processes during this period (however, see discussions below on the alternative possibility of energy-dependent transport). Finally, possible reconnection between the magnetic field lines enclosing the erupting flux rope and the ambient open field, as depicted in Figure~\ref{cartoon}(b), may provide a possible ``passway'' for the energetic particles to escape into the interplanetary space. This scenario is similar to those suggested by previous observational \citep{Maia2007, Demoulin2012} and modeling studies \citep{Masson2013, Masson2019}.

Energy-dependent transport through the interplanetary space could be an alternative scenario to account for the apparent energy-dependent release of energetic electrons observed by PSP. These electrons may experience various transport effects, such as particle scattering \citep{Strauss2020, Droge2000}, wave-particle interactions \citep{Kontar2009}, and cross-field transport \citep{Droge2016, Strauss2017}. In particular, \citet{Strauss2020} reported that higher-energy electrons ($\gtrsim$100 keV) may undergo stronger pitch-angle scattering effects than their lower-energy counterpart, which may potentially lead to the observed energy-dependent delay of \textit{in situ} electrons at higher energies. However, we argue that the agreement between the hardening of the HXR spectra and the \textit{in situ} energetic particle release times derived using different techniques generally favor the two-phase particle acceleration scenario.

Based on the velocity dispersion analysis of protons and electrons observed by PSP, which is closely aligned with the presumed Parker spiral connecting to the AR (with a separation of only 6$^{\circ}$; see Figure~\ref{Solar_mach}), the derived path lengths of the protons are about 22\% longer than the Parker spiral length. In comparison, the path length of energetic electrons derived from measurements made by the WIND and ACE spacecraft, which has a larger longitudinal separation from the AR ($26^{\circ}$), is 40\% longer than the nominal Parker spiral length. We attribute such excess particle path lengths to either the random walk of magnetic field lines \citep{Chhiber2021} or the diffusive particle transport in the interplanetary space \citep{Kouloumvakos2015, Malandraki2012}. Moreover, the relatively longer path length of particles arriving at the WIND/ACE spacecraft is a strong piece of evidence suggesting that cross-field diffusion might also play a role in the interplanetary transport of particles \citep{Strauss2017}. Such potential transport effects could lead to a longer decay in the observed flux profile and consequently extending the SEE duration observed by WIND/ACE (see, e.g., \citealt{Ruffolo1995}.) 

In our event, the ratio of the escaping \textit{in situ} electrons, based on ACE/WIND measurements, to HXR-emitting electrons near the solar surface is found to be extremely small, at the order of only 0.01--0.1\%, which seems lower than those reported previously \citep{Krucker1999, Wang2021}. In our case, as the WIND and ACE spacecraft had an appreciable longitudinal separation from the flare-hosting AR (27$^{\circ}$), cross-field diffusion might have contributed to the decrease of the peak electron flux that reached the spacecraft \citep{Strauss2020, Rodriguez2023}. 
On the other hand, understanding the origin of such a small population of escaping electrons remains an outstanding issue. \citet{Wang2021} suggested that the escaping \textit{in situ} electron population may arise from a postulated secondary acceleration site high in the corona. In our event, the existence (or non-existence) of such a secondary acceleration site is unclear. However, the close association of the release times of the \textit{in situ} electrons and protons with the flare impulsive phase suggests that it is more likely that they share the same origin as the microwave/HXR-emitting electrons. In this case, strong trapping or transport effects are required to limit the upward-propagating electron population to extremely small numbers. The recent study by \citet{Chen2024} provides a possible scenario with energetic particles trapped and accelerated in the above-the-loop-top magnetic bottle region under the conditions of strong diffusion. However, more observational and modeling studies are clearly required to understand such a stark departure from the equipartition of escaped electrons and those retained near the solar surface.

\begin{acknowledgements}
The authors thank S{\"a}m Krucker, William Setterberg, and Yixian Zhang for valuable discussions on HXR spectral analysis. 
This work is primarily funded by NASA HSO Connect grants 80NSSC20K1282 and 80NSSC20K1283 to NJIT (through a subcontract from SAO). M.W. and B.C. receive additional support from NSF SHINE grant AGS-2229338 to NJIT. H.W. is also support by NASA grant 80NSSC24M0174. The Expanded Owens Valley Solar Array (EOVSA) was designed and built and is now operated by the New Jersey Institute of Technology (NJIT) as a community facility. EOVSA operations are supported by NSF grant AGS-2130832 and NASA grant 80NSSC20K0026 to NJIT. The authors are grateful to the PSP/IS$\odot$IS, Fermi/GBM, SDO, e-Callisto, WIND, and ACE teams for making their data publicly available.
\end{acknowledgements}

\facilities{OVRO:SA, Fermi, Parker, WIND, ACE, SDO, GOES}

\vspace{5mm}

\bibliography{references}{}
\bibliographystyle{aasjournal}

\end{document}